# AC magnetic susceptibility study of a sigma-phase $Fe_{65.9}V_{34.1}$ alloy

M. Bałanda[1] and S. M. Dubiel[2*]

[1]The Henryk Niewodniczański Institute of Nuclear Physics, Polish Academy of Science, PL-31-342 Kraków, Poland, [2]AGH University of Science and Technology, Faculty of Physics and Applied Computer Science, PL-30-059 Kraków, Poland

## Abstract

A σ-phase $Fe_{65.9}V_{34.1}$ alloy was investigated with the AC magnetic susceptibility as a function of temperature, frequency and external magnetic field. An evidence was found that its magnetism shows features characteristic of a reentrant behavior viz. two transitions: first at $T_C$ ~312K from the paramagnetic state into the collinear ferromagnetic one, and second at $T_f$ ~302K to a mixed state (sometimes termed as a ferromagnetic re-entrant spin glass) which, finally, at a lower temperature ($T_{RSG}$ ~60K) transforms to a state where replica symmetry is broken. The frequency dependence of $T_f$ is lower than that of canonical spin glasses, a feature, that in the light of a high concentration of magnetic carriers, can be understood in terms of a weak coupling between magnetic clusters.

[*]Corresponding author: Stanislaw.Dubiel@fis.agh.edu.pl

## 1. Introduction

Sigma (σ) phase is a member of the Frank-Kasper (FK) family of phases also known as topologically close-packed (TCP) ones. Their characteristic feature are high values (12-16) of coordination numbers [1]. An interest in σ (and other FK-phases) has both practical as well as scientific reasons. The former follows from the fact that σ may precipitate in technologically important materials (steels) that – due to their excellent properties – have been used as construction materials in various branches of industry e. g. chemical, petrochemical, traditional and nuclear power plants etc. e. g. [2]. Once precipitated, it drastically deteriorates mechanical properties as well as decreases a corrosion resistance. Consequently, the interest in σ from the practical viewpoint is rather limited to a development of such steels in which the formation of σ does not occur or, at least, it is retarded. In turn, the scientific interest in σ follows from its complex crystallographic structure (tetragonal unit cell with 30 atoms distributed over 5 non-equivalent sublattices) - that is attractive in its own – and interesting physical properties that can be tailored by changing constituting elements and chemical composition (σ has not a definite stoichiometric composition). Concerning magnetic properties of the σ phase in binary alloys, the subject of the present study, only σ in Fe-Cr and Fe-V systems was known to be magnetic, and its magnetism was regarded as ferromagnetic [3,4]. Recently, the magnetism in the two systems was shown to be more complex, viz. it was found to be constituted by a spin glass (SG) of a re-entrant character [5]. Freshly, the magnetism of σ was discovered in Fe-Re [6] and in Fe-Mo [7,8] systems with SG being the ground state. The Fe-V system is especially interesting with regard to σ, as the phase can be formed within a wide range of composition, namely for ~33 - ~65 at% V [9]. This gives a unique chance for changing physical properties of σ. Concerning the magnetic ones, the magnetic ordering temperature (Curie point), $T_C$, can be continuously raised up above room temperature. The actual record obtained for the $Fe_{65.6}V_{34.4}$ alloy is as high as ~307K as determined from magnetization measurements or 324K as found from temperature dependence of the average hyperfine field [10]. In this paper we report results obtained with AC magnetic susceptibility techniques on a σ-$Fe_{65.9}V_{34.1}$ sample.

## 2. Experimental

### 2.1. Sample preparation and characterization

A master alloy of a nominal composition of $Fe_{66.5}V_{33.5}$ was prepared by an arc melting of appropriate amounts of elemental Fe (3N+ purity) and V (4N purity). The melting process was carried out in a protective atmosphere of pure argon and an ingot was flipped and re-melted three times to increase its homogeneity. The transformation of the α-phase master alloy into σ was performed by an isothermal annealing of the former in vacuum at 1273 K for 3 days followed by water quenching. Thus obtained sample was powdered and X-ray diffraction (XRD) pattern was recorded at room temperature – see Fig. 1. Its analysis with the program FullProf revealed that the original α-phase was transformed into σ to 97.5 % with a residual (2.5%) amount of Fe. Consequently, the stoichiometry of σ was $Fe_{59.9}V_{34.1}$. Details of the analysis will be given elsewhere [11].

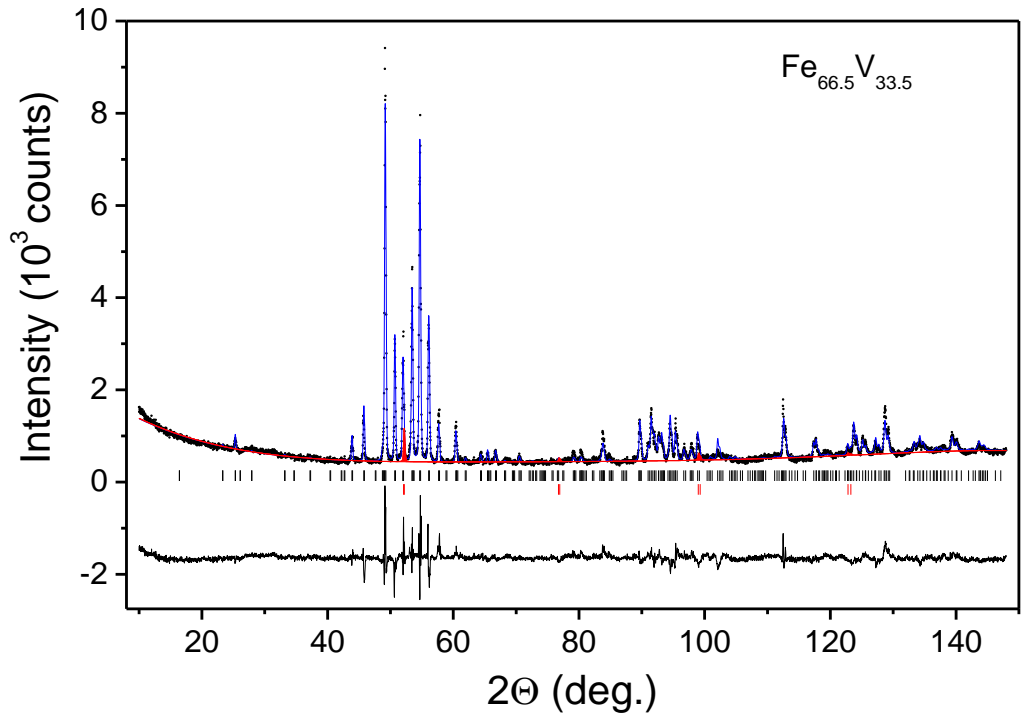

Fig. 1 The Rietveld refinement XRD pattern of the studied $Fe_{66.5}V_{33.5}$ sample recorded at room temperature. Positions of peaks of the σ and α phases are indicated by vertical bars in black and red, respectively. A difference pattern is also displayed.

### 2.2. Magnetic susceptibility measurements

AC magnetic susceptibility ($\chi_{AC}$) measurements were carried out by means of a MPMS SQUID magnetometer. The sample was in form of powder and it had a mass of 25.6 mg. Both real ($\chi'$) and imaginary ($\chi''$) components of $\chi_{AC} = \chi' - i\chi''$ were registered as a function of increasing temperature (T) in the range of 5-350 K. The frequency of the oscillating magnetic field from 1 Hz up to 1000 Hz was used and the dependence of the output signal on the amplitude of the driving field (1 Oe and 2 Oe) was tested. The influence of the applied DC magnetic field ($H_{DC}$) on the temperature behavior of $\chi_{AC}$ was investigated for $H_{DC}$ = 100, 200, 500 and 1000 Oe.
Temperature dependence of magnetization was measured in the constant field of 10 Oe in the ZFC and FC regime by means of a PPMS instrument.

### 3. Results and discussion

### 3.1. AC susceptibility in zero magnetic field

### 3.1.1. The maximum (cusp)

Real (in phase), $\chi'$, and imaginary (out of phase), $\chi''$, AC magnetic susceptibility components, measured as a function of temperature, $T$, and frequency, $f$, are shown in Fig. 2. The imaginary component amounts to 3% of the real one, which points to a rather weak relaxation effects in the studied alloy. Concerning $\chi'$, its shape is not typical of classical spin glasses (SGs), like for example $Cu$Mn, for which a cone-like curves with concave slopes and a well-defined casp were observed [12]. It should, however, be noticed that the typical shape of $\chi'$ was also measured for systems with high concentration of magnetic carriers and complex crystallographic structures e.g. for a $Tb_{117}Fe_{52}Ge_{113.8(1)}$ compound showing a cluster SG behavior [13], or for $\sigma$-$Fe_xMo_{100-x}$ alloys exhibiting several features characteristic of the canonical SGs despite high content of Fe viz. $43 \leq x \leq 53$ [7]. In contrast to $\chi'$, the $\chi''$ data are less regular. In our opinion, the interlacement of the $\chi''$ curves measured with different frequencies of the driving field, arises from the dissimilar anisotropies hence relaxation times of particular magnetic sub lattices.

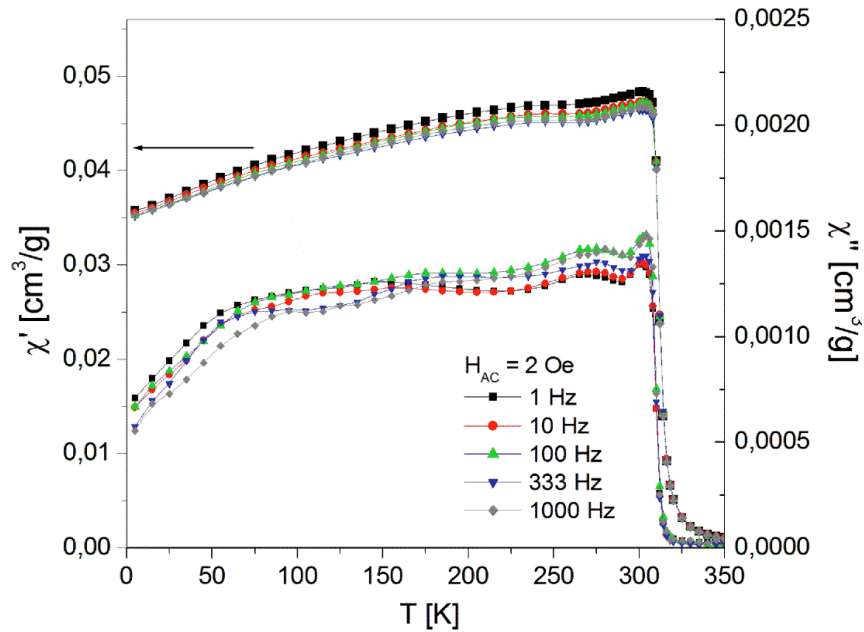

Fig. 2 Real, $\chi'$, and imaginary, $\chi''$, components of the AC magnetic susceptibility for $Fe_{59.9}V_{34.1}$ versus temperature for different frequencies indicated.

Our preliminary measurement of magnetization according to a field-cooled (FC) and zero-field-cooled (ZFC) protocol is shown in Fig. 3. For AC and DC measurements samples from the same batch but of the different masses were used. Fig. 3a clearly shows that the magnetization of the sample is irreversible. Moreover, the shape of the ZFC curve is virtually the same as that of $\chi'$ (see Fig.3b), as it occurs for soft magnetic materials. The ZFC-FC irreversibility is one of the characteristic features of different types of SGs, yet it is not unique to SGs.

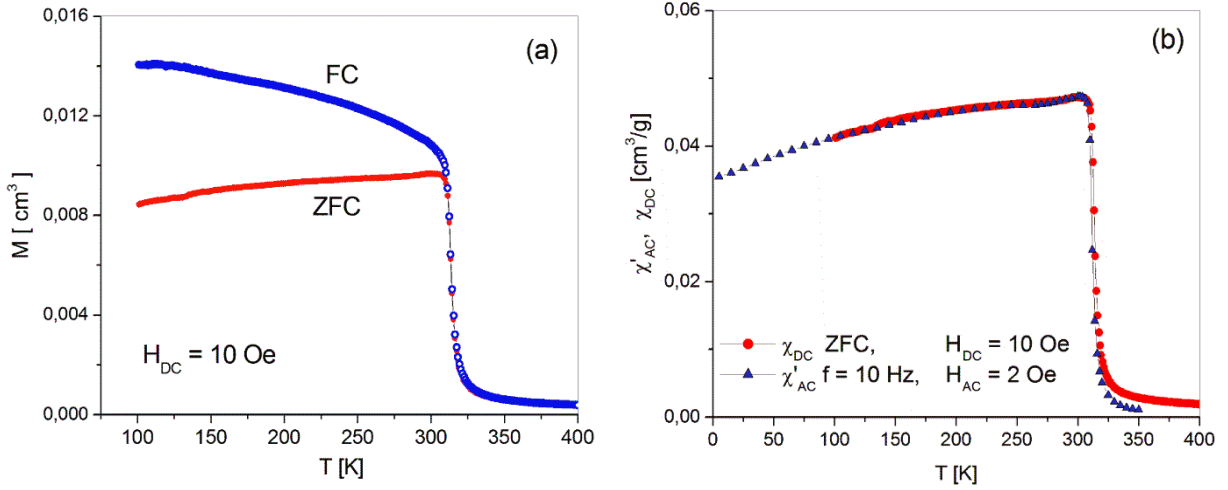

Fig. 3 (a) Temperature dependence of magnetization of the σ-$Fe_{65.9}V_{34.1}$ sample in zero-field-cooled (ZFC) and field-cooled (FC) regimes. (b) Direct comparison of AC and DC susceptibilities showing.

In order to shed more light on the nature of the irreversibility we will analyze a frequency dependence of characteristic temperatures in $\chi'$ and $\chi''$ curves in Fig. 2a. The most characteristic temperature, known as a spin freezing one, $T_f$, is the one at which the AC susceptibility has its maximum (cusp). Figure 2a gives evidence that this is the case both for $\chi'$ and $\chi''$. For spin glasses and superparamagnets (SPM) both the position of the maximum as well as its height depend on frequency, $f$. The former shifts towards higher temperature and the latter decreases. The shift of $T_f$, $\Delta T_f$, has been used to distinguish between SGs and SPMs. For this purpose the following figure of merit has been applied:

$$RST = \frac{\Delta T_f / T_f}{\Delta \log f} \qquad (1)$$

For SPMs the position of the cusp is more sensitive to frequency, hence the values of *RTS* are significantly higher than those for SGs. However, there is not well-defined border value of *RTS* between the two systems. As a rule of thumb, *RST* $\geq \sim 0.2$ is indicative of SPMs [14]. SGs can be further divided into two categories viz. canonical (paradigmatic, classical) ones like e.g. *Au*Mn, *Ag*Mn, *Au*Fe, *Cu*Mn for which *RTS* $\leq 0.08$ and cluster SGs e.g. with $0.08 < RST < 0.2$ [14]. Also here the border value of *RST* is not sharply defined, hence the classification into the two categories is not fully objective.

A presentation of $T_f$ data in terms of eq. (1) obtained from $\chi'$ curves for $H_{AC}$ = 1 Oe and 2 Oe is shown in Fig. 4. The dependence of $T_f$ on log$f$ determined for the $\chi''$ data was not linear for both $H_{AC}$ used.

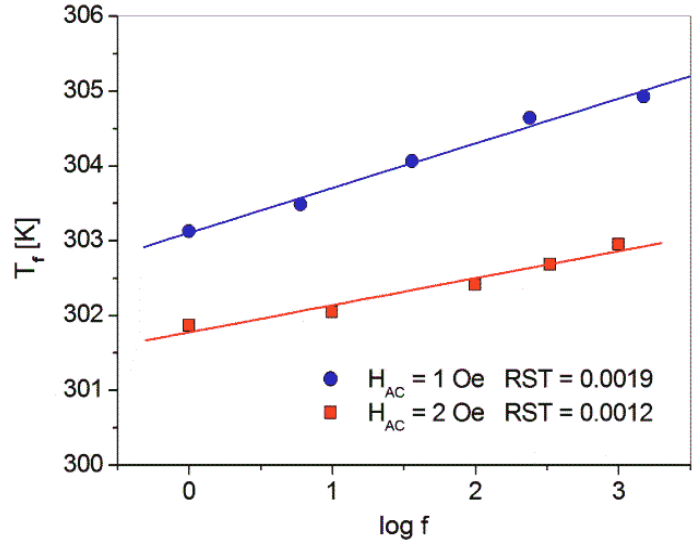

Fig. 4 Relationship between the $T_f$ and log$f$ for measurements carried out at two AC fields (1 and 2 Oe). The solid lines represent the linear fits in terms of eq. (1).

The *RST*-values derived therefrom slightly depend on the amplitude of the AC field, and are equal to 0.0019 and 0.0012 for $H_{AC}$ = 1 Oe and 2 Oe respectively. Such values are very small and even smaller than those found for the canonical SGs e. g. 005 for *Cu*Mn [12]. In terms of the above-given criterion, this means that spin glass features are only weakly manifested in the σ-Fe$_{65.9}$V$_{34.1}$ alloy. With reference to the mean-field theory of SGs, it testifies to very long-range interactions between magnetic entities in the sample. Noteworthy, no shift of $T_f$ we has been recently observed for a σ-phase Fe$_{54}$Cr$_{46}$ alloy [15] and that for σ-FeMo alloys was in the range of 0.0122-0.0135 [7].

Another question relevant to SGs is a description of the frequency dependence of the temperature of the maximum in the AC susceptibility, $T_f$. Three phenomenological laws can be applied in relation with this question: Arrhenius law – eq. (2), Vogel-Fulcher law – eq. (3) and the critical slowing-down law – eq. (4).

$$f = f_o \exp(-\frac{E_a}{k_B T_f}) \qquad (2)$$

$$f = f_0^{VF} \exp(-\frac{E_a}{k_B(T_f - T_0^{VF})}) \qquad (3)$$

$$f = f_o (\frac{T_f}{T_{SG}} - 1)^{zv} \qquad (4)$$

Here $k_B$ is the Boltzmann constant, $E_a$ activation energy, $T_o^{VF}$ the Vogel-Fulcher parameter, $T_{SG}$ a spin-glass temperature, and $zv$ is known as the dynamic exponent.

Application of eq. (2) resulted in non-physical values of fit parameters, in particular $f_o=10^{648}$ Hz and $E_a=1496$ $k_B$K. Such values cannot be regarded as physically meaningful. Interestingly, the application of the Arrhenius law to the *Cu*Mn system (Mn content less than ~6 at%) yielded the non-physical values for $E_a$ and $f_o$, as well [12]. On the other hand, the Vogel-Fulcher law yielded in our case reasonable values of the parameters viz. $f_o^{VF}=2\cdot10^{11}$ Hz, $E_a/k_B=101$K, $T_o^{VF}=298$K. Finally, by using the critical slowing-down law we obtained $T_{SG}=300.8$K and $zv=8.5$ which also looks correct.

The parameters obtained from the Vogel-Fulcher law have been used, via the Tholence criterion, to make a distinction between canonical and cluster SGs [16,17]. For this purpose one calculates the following quantity:

$$\delta T_f = \frac{T_f - T_o^{VF}}{T_f} \quad (5)$$

which in our case equals to 0.013, a value typical of the canonical SGs. For example $\delta T_f=0.07$ was found for *Cu*Mn [18]. In terms of an interaction between spin clusters, that may exist in our sample due to a high Fe content, such a small value of $\delta T_f$ as 0.013 may be also interpreted as indication of a very low degree of coupling between the clusters. As an alternative measure of the interactions between spin carriers freezing at $T_f$, hence that of the degree of magnetic clustering, has been used the ratio between the activation energy, $E_a$, and the Vogel-Fucher parameter $T_o^{VF}$ [19]. For the canonical SGs $E_a/k_B T_o^{VF} \approx 2$-$3$, whereas for cluster SGs the ratio should be much higher e. g. a value of 30 was reported for *LaAl$_2$*Gd [19]. In the present case the ratio is equal to 0.34, what again speaks in favor of a very weak coupling between the magnetic clusters, if they exist. The latter is plausible despite the high concentration of magnetic Fe atoms because the magnetism of σ-phase alloys is highly itinerant [6], hence the magnetic interactions have a very long-range dispersing thereby the inter-cluster magnetic coupling. On the other hand, the common interpretation of a non-zero value of $T_o^{VF}$ as an indication of the existence of magnetic clusters, and, via the Tholence criterion, as a measure of an inter-cluster interactions, may be put in doubt in the light of its "successful" application to "truly" canonical SGs like e. g. *Cu*Mn [12]. Clusters should be absent due to the low concentration of magnetic carriers (below ~6 at%). In such circumstances the frequency dependence of the spin-freezing temperature should be in line with the Arrhenius not with the Vogel-Fulcher law. However, the opposite was found to be true [12].

### 3.1.2. The magnetic ordering temperature

Previous experimental studies demonstrated that σ-Fe$_{100-x}$V$_x$ alloys order ferromagnetically [3,10]. The Curie temperature, $T_C$, strongly depends on the alloy composition ranging between ~15K for $x=55$ to ~315K for $x=34.5$ (in both cases the average over the values determined from the Mössbauer spectroscopic and DC magnetization measurements is given) [10]. In the light of these results it is reasonable to assume that also the presently studied sample orders ferromagnetically. Based on the previously found relationship between $T_C$ and $x$ [10], we can expect that the value of $T_C$ for the σ-Fe$_{66.5}$V$_{33.5}$ sample is the highest. We determined it here as the temperature of the inflection point in the χ′(T) curves measured in a low magnetic field (2 Oe) in different frequencies (1-1000 Hz). Data presented in Fig. 5 indicates that such-determined $T_C$ ≈312K is frequency independent what testifies to a true phase transition. It should be added that in the case of the σ-Fe$_{65.5}$V$_{34.5}$ sample $T_C$ ≈307K as determined from the inflection point in the magnetization curve [10], so the presently found value is the highest one, indeed, and in line with the trend revealed for other compositions [10].

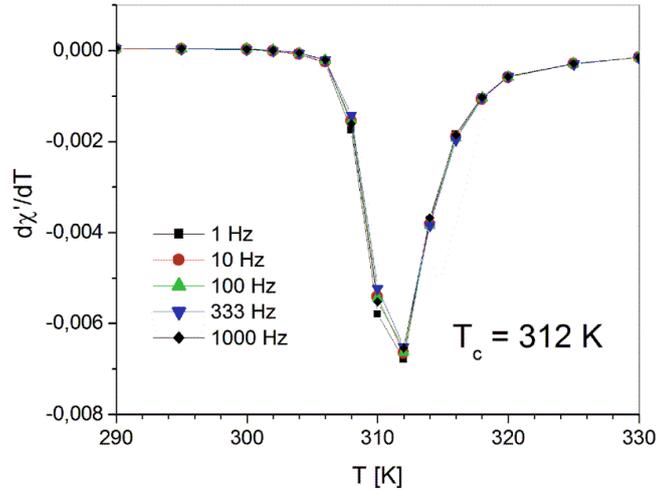

Fig. 5 $d\chi'/dT$ versus temperature in the temperature range where χ′ shows the fastest change. The minimum defines $T_C$.

**3.2. AC susceptibility in a biasing DC magnetic field**

An external magnetic field is known for its profound effect on the AC susceptibility of SGs. Especially sensitive is the temperature region in the vicinity of $T_f$, a feature that was already revealed in the early days of SGs [20]. So it is of interest to investigate such effect to further explore SGs. As illustrated in Figs. 6 and 7, our measurements are in line with the above-mentioned statement. Concerning χ′, in rather small fields (100 and 200 Oe) the characteristic cusp at $T_f$ has been severely rounded and shifted towards lower temperature while in stronger fields (500 and 1000 Oe) a significant suppression can be observed accompanied by an appearance of a small peak in the vicinity of $T_f$. Its exact position is situated at ~313K and ~315K for $H_{DC}$=500 and 1000 Oe, respectively, hence close to the Curie temperature of ~312K as determined from the inflection point in χ′ measured in zero DC field. A similar behavior of χ′ was observed in some other systems showing reentrant SGs e. g. in NiCoMnSb Heusler alloys [21], and, interestingly, in the ferromagnetic gadolinium [22], the disordered Cr$_3$Fe ferromagnet [23] or in the layered 2D Heisenberg magnet [24]. Such an anomaly has been explained in terms of short range fluctuations close the order transition. Regarding the rounded part of χ′ there is a striking difference between the curves recorded in the fields of 100 and 200 Oe and those measured at 500 and 1000 Oe. Namely, in the former ones one observes its downward inclination in the low temperature range, however, the temperature at which this trend begins cannot be precisely defined. It is likely related with a transition into a SG "phase" where the transverse components of spins are frozen and replica symmetry is spontaneously broken [25]. The existence of such transition located at ~60K, known as the re-entrant temperature, $T_{RSG}$, can be better seen in the $\chi''$ curves – Fig. 6 where a steep decrease of $\chi''$ occurs for T <~60K. The latter can be taken as an indication of a strong departure from the collinear structure, and a strong irreversibility phenomena in the related state.

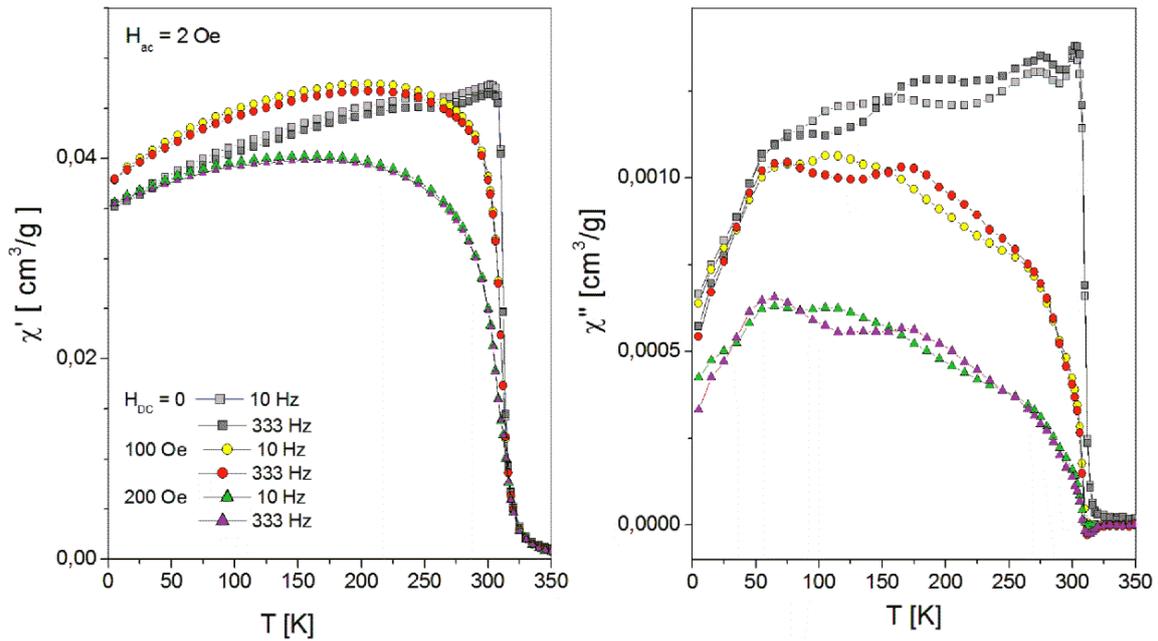

Fig. 6 Real, $\chi'$, and imaginary, $\chi''$, parts of the AC susceptibility versus temperature, recorded for the σ-$Fe_{66.5}V_{33.5}$ sample in a biasing DC magnetic field of 100 Oe and 200 Oe for $H_{ac}$ = 2 Oe and two frequencies shown. Data obtained in zero bias field are added for comparison.

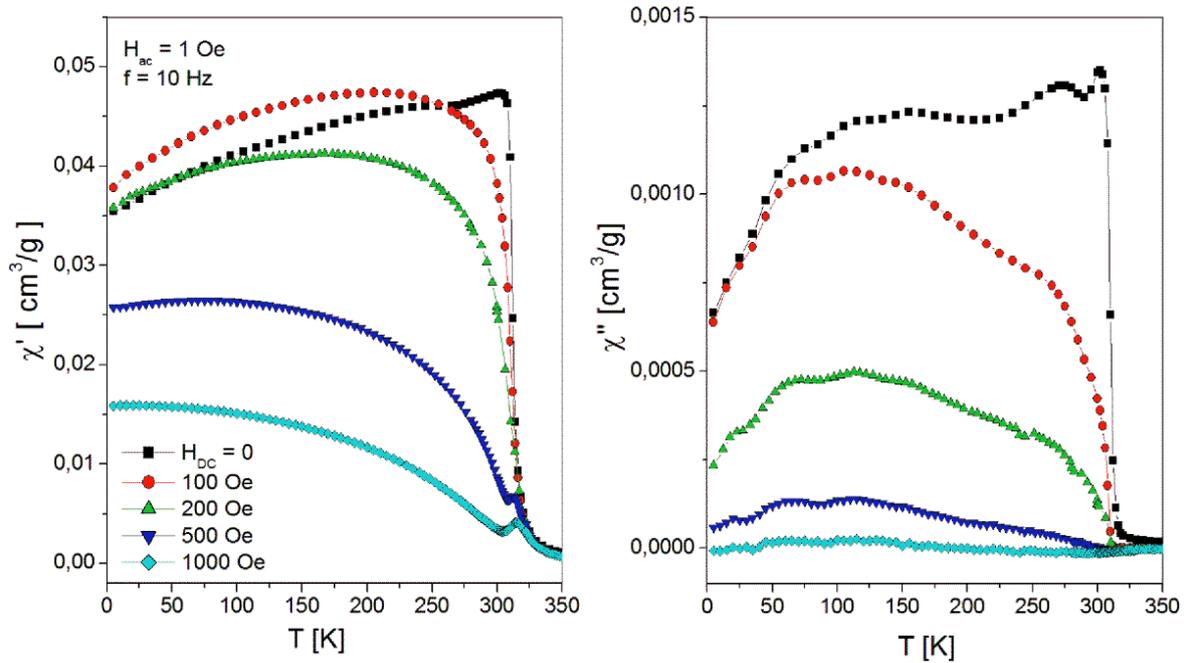

Fig. 7 The effect of a biasing DC magnetic field in the range 100 - 1000 Oe on $\chi'$ and $\chi''$ for σ-$Fe_{66.5}V_{33.5}$ measured at $H_{ac}$ = 1Oe and f = 10 Hz.

## 4. Summary

AC magnetic susceptibility measurements versus temperature (4-350 K), frequency (1-1000 Hz) and external magnetic field (100-1000 Oe) were carried out on a sigma-phase $Fe_{65.9}V_{34.1}$ alloy. They revealed that the magnetism of the investigated sample has a reentrant character: at ~312K the ferromagnetic order sets in, and at ~302K there is a transition into a spin-glass-like state which apparently is mixed i.e. the spin glass coexists with the ferromagnetic ordering. At a still lower temperature viz. ~60K a transition into the ground state was found. This scenario seems to be in line with the mean-field model introduced by Gabay and Thoulouse [25]. Interestingly, all figures of merit used to characterize SGs like the Tholence criterion, shift of the spin-freezing temperature per decade of frequency, the ratio between the activation energy and the Vogel-Fulcher temperature have values lower than these for canonical spin glasses. This implies that a magnetic coupling between spin clusters, whose existence in the investigated system is likely due to the high concentration of Fe, is very weak. The weakness of the coupling may result from the itinerant character of magnetism observed in the σ-phase systems, as long-range magnetic interactions are expected to diffuse inter cluster interactions.

**Acknowledgement**

J. Żukrowski is thanked for making the master alloy and J. Przewoźnik for the XRD analysis.

**References**

[1] A. K. Sinha, *Topologically Close-packed Structures of Transition Metal Alloys,* Pergamon Press Ltd., Oxford, 1972

[2] M. Venkatraman, K. Pavitra, V. Jana, T. Kachwala, Adv. Mater. Res., 794 (2013) 163

[3] D. Parsons, Nature, 185 (1960) 839

[4] D. A. Read, E. H. Thomas, J. B. Forsythe, J. Phys. Chem. Solids, 29 (1968) 1569

[5] R. Barco, P. Pureur, G. L. F. Fraga, S. M. Dubiel, J. Phys.: Condens. Matter, 24 (2012) 046002

[6] J. Cieślak, S. M. Dubiel, M. Reisner, J. Toboła, J. Appl. Phys, 116 (2014) 183902

[7] J. Przewoźnik, S. M. Dubiel, J. Alloy. Comp., 630 (2015) 222

[8] J. Cieślak, S. M. Dubiel, M. Reissner, arXiv:1411.2446 (2014)

[9] http://www.calphad.com/iron-vanadium.html

[10] J. Cieślak, B. F. O. Costa, S. M. Dubiel, M. Reissner, W. Steiner, J. Magn. Magn. Mater., 321 (2009) 2160

[11] J. Żukrowski, S. M. Dubiel, to be published

[12] C. A. M. Mulder, A. J. van Duyneweldt, J. A. Mydosh, Phys. Rev. B, 23 (1981) 1384

[13] J. Liu, W. Xie, K. A. Gschneidner Jr, G. J. Miller, V. K. Pecharsky, J. Phys.: Condens. Matter, 26 (2014) 416003

[14] J. A. Mydosh, *Spin glasses: An experimental introduction*, vol. 125, Taylor & Francis, London, 1993

[15] S. M. Dubiel, M. I. Tsindlekht, I. Felner, J. Alloy. Comp., 642 (2015) 177

[16] J. L. Tholence, Phys. B 126 (1984) 157

[17] Y. Yeshurun, J. L. Tholence, J. K. Kjems, B. Wanklyn, J. Phys. C, 18 (1985) L 483

[18] S. Shtrikman, E. P. Wohlfarth, Phys. Lett. A, 85 (1981) 467

[19] D. Fiorani, J. L. Tholence, J. L. Dormann, J. Phys. C, 19 (1986) 5495

[20] V. Cannela, J. A. Mydosh, Phys. Rev. B, 6 (1971) 4220

[21] Ajaya K. Nayak, K. G. Suresh, A. K. Nigam, J. Phys.: Condens. Matter, 23 (2011) 416004

[22] S. Yu. Dan'kov, A. M. Tishin, V. K. Pecharsky, K. A. Gschneidner, Jr., Phys. Rev. B, 57 (1998) 3478

[23] S.F. Fisher, S.N. Kaul, H. Kronmüller, J. Magn. Magn. Mater. 272-276 (2004) 254.

[24] C.C. Becerra, A. Paduan-Filho, Solid State Commun. 125 (2003) 99.

[25] M. Gabay, G. Toulouse, Phys. Rev. Lett., 56 (1981) 201